\documentclass[footinbib,a4paper,aps,superscriptaddress,twocolumn,preprintnumbers,amsmath,amssymb,nobalancelastpage,10pt,prl]{revtex4-2}
\usepackage{amsmath}
\usepackage{amssymb}
\usepackage{tikz}
\usetikzlibrary{shapes}
\usepackage{verbatim}
\usepackage{graphicx}
\usepackage{color}
\usepackage{braket}
\usepackage{yfonts}
\usepackage{soul}
\usepackage[normalem]{ulem}
\usepackage{dsfont}
\usepackage[colorlinks=true,citecolor=blue,linkcolor=blue,urlcolor=blue]{hyperref}
\usepackage{bm}

\begin{document}
 
\title{Observation of string breaking on a ($2+1$)D Rydberg quantum simulator}

\author{Daniel Gonz\'alez-Cuadra}
\affiliation{Institute for Theoretical Physics, University of Innsbruck, 6020 Innsbruck, Austria}
\affiliation{Institute for Quantum Optics and Quantum Information of the Austrian Academy of Sciences,
6020 Innsbruck, Austria}

\author{Majd Hamdan}
\affiliation{QuEra Computing Inc., 1284 Soldiers Field Road, Boston, MA, 02135, USA}

\author{Torsten V. Zache}
\affiliation{Institute for Theoretical Physics, University of Innsbruck, 6020 Innsbruck, Austria}
\affiliation{Institute for Quantum Optics and Quantum Information of the Austrian Academy of Sciences,
6020 Innsbruck, Austria}

\author{Boris Braverman}
\affiliation{Department of Physics, University of Toronto, 60 St. George Street, Toronto, ON M5S 1A7, Canada}
\affiliation{QuEra Computing Inc., 1284 Soldiers Field Road, Boston, MA, 02135, USA}

\author{Milan Kornjača}
\affiliation{QuEra Computing Inc., 1284 Soldiers Field Road, Boston, MA, 02135, USA}

\author{Alexander Lukin}
\affiliation{QuEra Computing Inc., 1284 Soldiers Field Road, Boston, MA, 02135, USA}

\author{Sergio H. Cant\'u}
\affiliation{QuEra Computing Inc., 1284 Soldiers Field Road, Boston, MA, 02135, USA}

\author{Fangli Liu}
\affiliation{QuEra Computing Inc., 1284 Soldiers Field Road, Boston, MA, 02135, USA}

\author{Sheng-Tao Wang}
\affiliation{QuEra Computing Inc., 1284 Soldiers Field Road, Boston, MA, 02135, USA}

\author{Alexander Keesling}
\affiliation{QuEra Computing Inc., 1284 Soldiers Field Road, Boston, MA, 02135, USA}

\author{Mikhail D. Lukin}
\affiliation{Department of Physics, Harvard University, Cambridge, MA 02138, USA}

\author{Peter Zoller}
\affiliation{Institute for Theoretical Physics, University of Innsbruck, 6020 Innsbruck, Austria}
\affiliation{Institute for Quantum Optics and Quantum Information of the Austrian Academy of Sciences,
6020 Innsbruck, Austria}

\author{Alexei Bylinskii}
\affiliation{QuEra Computing Inc., 1284 Soldiers Field Road, Boston, MA, 02135, USA}

\begin{abstract}

Lattice gauge theories (LGTs) describe a broad range of phenomena in condensed matter and particle physics. A prominent example is confinement, responsible for bounding quarks inside hadrons such as protons or neutrons~\cite{gross202350}. When quark-antiquark pairs are separated, the energy stored in the string of gluon fields connecting them grows linearly with their distance, until there is enough energy to create new pairs from the vacuum and break the string. While such phenomena are ubiquitous in LGTs, simulating the resulting dynamics is a challenging task~\cite{Bauer2023}. Here, we report the observation of string breaking in synthetic quantum matter using a programmable quantum simulator based on neutral atom arrays~\cite{Ebadi_2021,Scholl_2021,Wurtz_2023}. We show that a (2+1)D LGT with dynamical matter can be efficiently implemented when the atoms are placed on a Kagome geometry~\cite{Samajdar_2021}, with a local U($1$) symmetry emerging from the Rydberg blockade~\cite{Surace_2020}, while long-range Rydberg interactions naturally give rise to a linear confining potential for a pair of charges, allowing us to tune both their masses as well as the string tension. We experimentally map out the corresponding phase diagram by adiabatically preparing the ground state of the atom array in the presence of defects, and observe substructure of the confined phase, distinguishing regions dominated by fluctuating strings or by broken string configurations. Finally, by harnessing local control over the atomic detuning, we quench string states and observe string breaking dynamics exhibiting a many-body resonance phenomenon. Our work paves a way to explore phenomena in high-energy physics using programmable quantum simulators.

\end{abstract}

\maketitle

\renewcommand{\figurename}{\textbf{Fig.}}

\begin{figure*}
    \centering
    \includegraphics[width=1.0\linewidth]{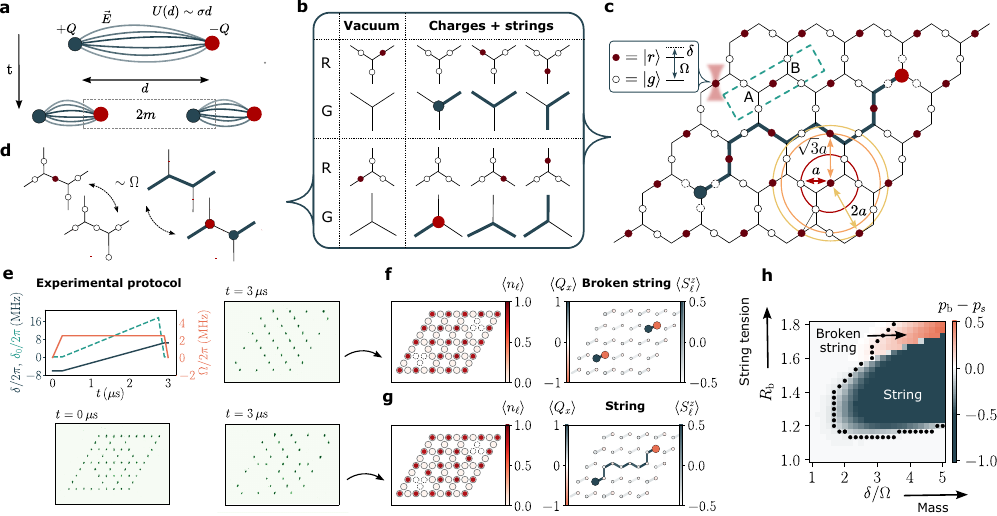}
    \caption{{\bf Emergent confinement and string breaking on a ($2+1$)D Rydberg atom array:} (a) U(1) gauge theories satisfy local Gauss's law constraints, where electric charges $\pm Q$ are connected by strings of electric field $\vec{E}$. In a confined phase, the potential energy $U(d)$ between charges increases linearly with their distance $d$, until the strings break by creating particle pairs with masses $2m$. (b) Equivalence between atom configurations satisfying the Rydberg blockade constraint (R) and the corresponding gauge-invariant states (G), shown for the A (upper) and B (lower) sites in the unit cell of (c). For the vacuum $S^z_\ell = -1/2$ (black lines) $\forall \ell$, while defects correspond to $Q_x=\pm 1$ charges (blue/red circles) and $S^z_\ell = +1/2$ strings (blue lines). (c) Neutral atoms trapped in optical tweezers are arranged on the links of a hexagonal lattice, with a two-site unit cell (green) and lattice spacing $a$. The figure depicts a Rydberg ordered configuration / LGT vacuum, with a line defect / string connecting two static charges, forced by removing atoms from the array (dashed circles). Rydberg atoms in the string interact with their second neighbors, giving rise to a confining potential with string tension $\sigma\approx V(\sqrt{3}a) - V(2a)$. (d) String breaking in the Rydberg system, where particle-pairs are created by flipping the state of one atom. (e) State preparation protocol involving local detuning, where we show experimental snapshots of the atom array in the initial disorder phase, and after deterministically preparing a broken string state (top)  or one of the 6 degenerate unbroken string states (bottom). Dark spots correspond to atoms detected in the $\ket{g}$ state. 
    (f) and (g) show the real-space configurations of these two prepared states, obtained by collecting these snapshots (results for the remaining 5 string states for this placement of charges are shown in Extended Data Fig.\ref{fig:figm1}). We depict the expectation value of the Rydberg densities (left), and the corresponding LGT observables (right). (h) Phase diagram calculated theoretically for the Rydberg Hamiltonian~\eqref{eq:Rydberg_H} as a function of $R_{\rm b}$ and $\delta / \Omega$, for the geometry displayed in (c). We plot the difference $p_{\rm b} - p_{\rm s}$ between the probability of the broken $p_{\rm b}$ and the unbroken string configurations $p_{\rm s}$ in the ground state, showing two distinct regions within the ordered phase found in \cite{Samajdar_2021}, marked here by black dots (see Methods).}
    \label{fig:fig1}
\end{figure*}

The observation that certain fundamental particles such as quarks cannot be isolated but are instead bound in composite hadrons like protons or neutrons is attributed to a phenomenon called \emph{confinement}. 
According to quantum chromodynamics (QCD)~\cite{gross202350}, an extensive amount of energy is stored in a flux string of gluon fields that confines, say, a quark-anti-quark pair. 
Upon separating two quarks, this linearly confining potential provides the energy to create new particle pairs, thereby breaking a long flux string into shorter ones [see Fig.~\ref{fig:fig1}(a)]. 
When quarks are produced in high-energy particle colliders, they are thus detected indirectly as this process of \emph{string breaking} leads to ``jets'' of secondary particles~\cite{gross202350}.

Related phenomena are ubiquitous in all gauge theories reminiscent of QCD, including \emph{lattice} models with local (gauge) symmetry constraints~\cite{Hebenstreit_2013, Kuhn_2015, Pichler_2016, Verdel_2023}.
Such lattice gauge theories (LGTs)~\cite{Montvay_1997} also emerge in condensed matter physics, e.g., in the context of topologically ordered quantum matter~\cite{Sachdev_2019} or deconfined quantum critical points~\cite{Senthil_2004}. Recent progress in creating, manipulating and imaging quantum systems based on neutral atoms, trapped ions and superconducting qubits offers the possibility to realize \textit{synthetic quantum matter} with tunable interactions, and to use them as analog quantum simulators~\cite{Altman_2021}.
Realizing synthetic LGTs in such systems enables the \emph{direct exploration} of confinement and string breaking dynamics with high spatio-temporal resolution~\cite{Banerjee_2012, Wiese_2013, Zohar_2015, Banuls_2020, Aidelsburger_2022, Klco_2022, Di_Meglio_2024}. We note that the  calculation of such real-time evolution is generally believed to be classically intractable at large scales~\cite{Bauer2023}, making  quantum simulation experiments especially valuable. 

Here we report an observation of string breaking in
two-dimensional synthetic quantum matter. We employ
a programmable Rydberg atom array operating in an
analog quantum simulator mode~\cite{chen2023continuous,shaw2024benchmarking,Manovitz_2024,anand2024dual}. In contrast to
complementary on-going efforts to explore real-time LGT
dynamics with digital quantum computers~\cite{Google_2024}, string
breaking emerges in our experiment as a many-body phenomenon, based on a mapping of the physically realized Hamiltonian to a confining U(1) LGT with
dynamical matter. While previous implementations of
LGTs have focused on (1+1)D~\cite{Martinez_2016, Kclo_2018, Schweizer_2019, Kokail_2019, Mil_2020, Yang_2020, Zhou_2021, Nguyen_2021, Tan_2021, Frolian_2022, Mildenberger_2022, Meth_2023, De2024}, here we realize
two-dimensional geometries, allowing for transversally fluctuating strings. This enables mapping an equilibrium
phase diagram of broken and unbroken strings in ($2+1$)D, using a combination of local control and quasi-adiabatic state
preparation. We further observe the breaking of initially prepared string configurations
in real-time dynamics, following the quench of a local control parameter.

\section*{String breaking in synthetic quantum matter}

\paragraph{Key elements of U(1) gauge theories.}
The defining feature of a gauge theory is its invariance under local transformations, which implies a local conservation law. In the case of a U(1) gauge theory such as quantum electrodynamics (QED), this constraint takes the form of the familiar Gauss's law, which relates the charge density $\rho$ at a given point in space with the divergence of the electric field $\vec{E}$: $\nabla \cdot \vec{E} - \rho = 0$ [Fig.~\ref{fig:fig1}(a)].

Similar local constraints are present in all U($1$) gauge theories, including their lattice versions, where the physical Hilbert space of the theory includes states where pairs of charges are connected by electric fluxes (strings)~\cite{Kogut_1975}. The prototypical LGT Hamiltonian contains both (pure gauge) electric field terms $E^2_\ell$, assigning an energy to different string configurations, as well gauge-matter interactions, $a^\dagger_x e^{iA_\ell} a^{\vphantom{\dagger}}_y$, which create pairs of charges at neighboring sites $x$ and $y$ connected by strings on links $\ell = \langle x, y \rangle$. Here $e^{iA_\ell}$ is a lattice version of the parallel transporter, with $[E_\ell, e^{iA_{\ell^\prime}}] = \delta_{\ell, \ell^\prime} e^{iA_\ell}$, and $a^{(\dagger)}$ represent fermionic (as in QED) or bosonic particles (as considered here).

For LGTs in a confining phase (typically driven by the $E^2_\ell$ term), the energy associated to a pair of particles increases with their separation $d$, where a linear dependence, $U(d) \sim \sigma d$, defines the string tension $\sigma$. For sufficiently large $d$, this energy can produce new particle pairs from the vacuum, thus breaking the string that connects the charges [Fig.~\ref{fig:fig1}(a)]. Observing string breaking in synthetic quantum matter therefore requires implementing local Gauss's law constraints as well as a confining potential between charge excitations, and below we show how both ingredients are naturally available in Rydberg atom arrays. 

\paragraph{From a Rydberg atom array to a U(1) LGT.}
Our experiments employ a programmable 
tweezer array trapping ${}^{87}$Rb atoms on the sites of a two-dimensional Kagome lattice, or equivalently the links $\ell$ of a hexagonal lattice [see Fig.~\ref{fig:fig1}(c) and Methods].
Starting from a ground state $\ket{g}$, the atoms are laser-coupled to a Rydberg state $\ket{r}$ with Rabi frequency $\Omega$ and detuning $\delta$. 
Including native van der Waals interaction $V_{\ell,\ell^\prime} = C_6 / (\|{\bf x}_\ell - {\bf x}_{\ell^\prime}\|/a)^6$ with constant $C_6$ and $a$ the lattice spacing, our system can be described by the Hamiltonian
\begin{equation}
\label{eq:Rydberg_H}
\begin{aligned}
H_{\rm Ryd} & = \frac{\Omega}{2}\sum_\ell X_\ell - \delta \sum_\ell n_\ell+ \frac{1}{2}\sum_{(\ell, \ell^\prime)} V_{\ell,\ell^\prime} n_\ell n_{\ell^\prime} \;,
\end{aligned}
\end{equation}
with $X_\ell = \ket{g}_\ell\!\bra{r} + {\rm H.c.}$ and $n_\ell = \ket{r}_\ell\!\bra{r}$, and we set $\hbar = a = 1$ in the following.

The key idea of our approach is to use 
strong interactions to implement a local constraint that can be recast as a U(1) Gauss's law~\cite{Surace_2020}. Here, we are specifically interested in a confined phase at large $\delta/\Omega$ and $1.2 \lesssim R_{\rm b} \lesssim 1.8$~\cite{Samajdar_2021}, where $R_{\rm b} = (C_6 / 2\Omega)^{1/6}$ is the blockade radius, associated with so-called nematic phase \cite{Samajdar_2021}. As we show below, both confined charges and strings emerge as low-energy excitations on this phase [Fig.~\ref{fig:fig1}(c)].
More precisely, in the Rydberg blockade regime ($R_{\rm b}^6 \gg \delta/\Omega$) at most one atom within the blockade radius can be excited to a Rydberg state.
In our chosen geometry, this subspace contains only four configurations in the neighborhood of each site [Fig.~\ref{fig:fig1}(b)]. To map a state $\ket{\psi}$ of the atoms in this subspace to a LGT configuration, we introduce U($1$) gauge fields as spin-$1/2$ operators for every link $\ell$, satisfying $[S^z_\ell, S^{\pm}_{\ell^\prime}] =\pm S^{\pm}_\ell\delta_{\ell,\ell^\prime}$ and $[S^+_\ell,S^-_{\ell^\prime}] = 2S^z_\ell\delta_{\ell,\ell^\prime}$, and additionally matter fields as hardcore-boson operators $a^{(\dagger)}_x$, with $[a^{\vphantom{\dagger}}_x, a^\dagger_{y}] = \delta_{x,y}$ and $a^\dagger_xa_x \in \{0,1\}$, associated to each site $x$.
 As discussed in detail in the Methods, every allowed state $\ket{\psi}$ fulfills the constraint $G_x\ket{\psi} = q_x\ket{\psi}$, where
\begin{equation}
\label{eq:gauge_generators}
    G_x = \nabla_x S^z - Q_x \;
\end{equation}
are the local generators of a U(1) gauge symmetry, with $\nabla_x S^z \equiv (-1)^{s_x}\sum_{\ell\in x} (-1)^{s_\ell} S^z_\ell$. Here, $Q_x = a^\dagger_x a^{\vphantom{\dagger}}_{x} - [1 - (-1)^{s_x}]/2$ denotes dynamical charges, and without defects our experimental setup corresponds to the staggered static charge configuration $q_x = (-1)^{s_x}/2$. (Note that $s_x$ and $s_\ell$ are spatially varying signs defined in Methods).

As a result, in the blockade regime, our system can be described by the following LGT Hamiltonian,
\begin{equation}
\label{eq:LGT_H}
\begin{aligned}
    &H_{\rm eff} = \frac{\Omega}{2} \sum_{\langle x,y\rangle} \left[a^\dagger_x S^+_{\langle x,y \rangle}a^{\vphantom{\dagger}}_y + \text{H.c.}\right] + \frac{\delta}{2}\sum_x (-1)^{s_x} a^\dagger_x a^{\vphantom{\dagger}}_x \\
    & + \frac{1}{2}\sum_{(\ell, \ell^\prime)\neq \langle\ell, \ell^\prime\rangle}V_{\ell,\ell^\prime}\left[(-1)^{s_\ell}S^z_\ell + \frac{1}{2}\right]\left[(-1)^{s_{\ell^\prime}}S^z_{\ell^\prime} + \frac{1}{2}\right] \;,
\end{aligned}
\end{equation}
and below we will see that this is approximately true in a larger region away from this limit.
In contrast to standard QED, our system corresponds to a variant of so-called quantum link models~\cite{Wiese_2013} where the usual field content is truncated while maintaining an exact local U($1$) symmetry, i.e. $\left[H_{\rm eff}, G_x\right] = 0$.  In particular, the electric field $E_\ell$ takes only two discrete values, represented by spin-1/2 operators $S^z_\ell$, and the link operators $e^{iA_\ell}$ are approximated by $S^+_\ell$. 

The first term $\propto \Omega>0$ in Eq.~\eqref{eq:LGT_H} accounts for gauge-matter interactions, creating and destroying particle pairs [Fig.~\ref{fig:fig1}(d)], while the second one $\propto \delta \equiv 2m_0$ gives a bare mass $m_0$ to the matter degrees of freedom. A unique feature of our setup is the presence of long-range interactions $\propto V_{\ell,\ell^\prime}$ that assign an electric energy to different gauge field configurations, where we identify $S^z_\ell = +\frac{1}{2}$ as strings. This term plays the role of $E_\ell^2$ and gives rise to a linearly confining potential with string tension $\sigma \propto R_\text{b}^6$ (see Fig.~\ref{fig:fig1}(c) and Methods for details).

Positive and negative static charges can be realized in our setup by removing three neighboring atoms in respective unit cells from the array [Fig.~\ref{fig:fig1}(c)].  In Fig.~\ref{fig:fig1}(f) and Fig.~\ref{fig:fig1}(g), we show approximated product states corresponding to broken and unbroken strings connecting these charges, respectively, prepared experimentally in the Rydberg atom array using a state preparation protocol involving local detuning control (see Fig.~\ref{fig:fig1}(e) and Methods). We note that, although these product states correspond in general to excited states of the model \eqref{eq:LGT_H}, we expect the ground state to transition between these unbroken and broken string configurations as we increase the string tension (controlled by $R_{\rm b}$) or decrease the mass (controlled by $\delta / \Omega$), as shown in Fig.~\ref{fig:fig1}(h) for a fixed distance $d$ (measured in unit cells). In the following, we employ the Rydberg atom array to investigate this string breaking phase diagram experimentally.

\begin{figure*}
    \centering
    \includegraphics[width=1.0\linewidth]{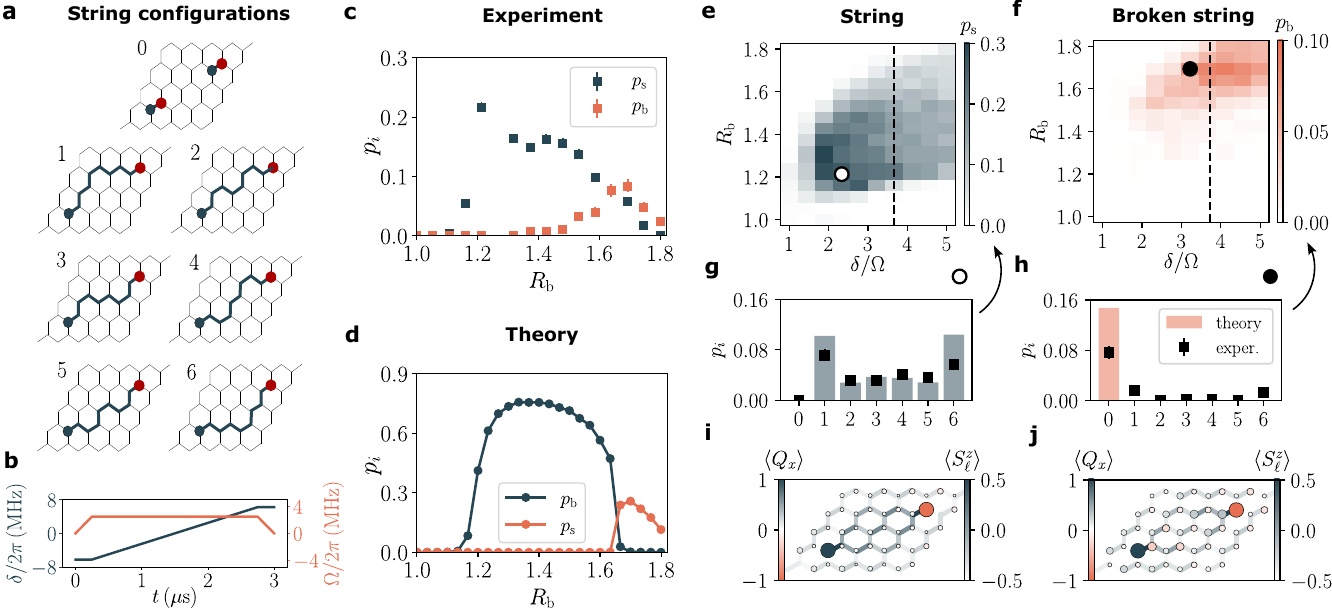}
    \caption{{\bf String breaking in equilibrium} (a) Product state configurations corresponding to a broken string (0) as well as to unbroken strings of the same (minimal) length (1--6) connecting two charges displaced in both directions, $d_0 = d_1 = 2$, for a system with $L_0 = L_1 = 5$. (b) Quasi-adiabatic state preparation protocol (details in Methods), where the final value of $\delta$ is varied to scan along the x-axis in the phase diagram. (c) Values of $p_{\rm b}$ and $p_{\rm s}$ as a function of $R_{\rm b}$ and fixed $\delta / \Omega = 3.67$ [dashed line in (e) and (f)] obtained experimentally by quasi-adiabatically preparing the ground state of the Rydberg atom array. Here $p_{\rm s}$ grows from zero to a finite value in the string region, while it decreases again in the broken string region. At the same time, $p_{\rm p}$ acquires a finite value in the latter, indicating string breaking, consistent with the theoretical prediction computed numerically (d). The coexistence of broken and unbroken configurations in the experiment is likely due to imperfect adiabatic state preparation, since each of them is a low-energy excited state in the region dominated by the other one. (e) and (f) show the values of $p_{\rm s}$ and $p_{\rm b}$ obtained experimentally as a function of $R_{\rm b}$ and $\delta / \Omega$, respectively. (g) and (h) show the distribution of probabilities $p_i$, where $i=0,...,6$ correspond to the classical configurations shown in (a), for $R_{\rm b} = 1.2$ and $\delta / \Omega = 2.3$ (white circle) and  $R_{\rm b} = 1.7$ and $\delta / \Omega = 3.2$ (black circle), respectively, both for theory and experiment. (i) and (j) show the corresponding real-space configurations obtained in the experiment. 
    }
    \label{fig:fig3}
\end{figure*}

\section*{String breaking phase diagram}

We leverage the two-dimensional nature of our programmable quantum simulator ``Aquila'' \cite{Wurtz_2023} and study string breaking of  fluctuating ($2+1$)D strings. For this, we consider a system with $L_0 = L_1 = 5$ unit cells, where $0$/$1$ denote the two axes of the hexagonal lattice, and place static charges separated in both directions, with $d_0 = d_1 =2$, for a total of $59$ atoms [same as in Fig.~\ref{fig:fig1}(c)]. Other geometries are presented in Methods. As depicted in Fig.~\ref{fig:fig3}(a), there are six different string configurations with the same (minimal) length connecting these two static charges, and the ground state in the unbroken regime is thus expected to consist of a superposition of them. To prepare the ground state of the Rydberg Hamiltonian~\eqref{eq:Rydberg_H} experimentally, we implement an quasi-adiabatic state-preparation protocol (details in Methods). As depicted in Fig.~\ref{fig:fig3}(b), we start from the disordered phase at a large negative detuning and at a fixed value of $R_{\rm b}$. We then apply $\Omega$ globally and sweep $\delta$, corresponding to an evolution that, when perfectly adiabatic, ends with the ground state of Eq.~\eqref{eq:Rydberg_H} at a finite positive value of $\delta / \Omega$. As this protocol crosses a symmetry-breaking phase transition, where the gap closes as a function of the system size, the finite rate of sweep introdcues defects to the target ground state. 

\begin{figure}
    \centering
    \includegraphics[width=1.0\linewidth]{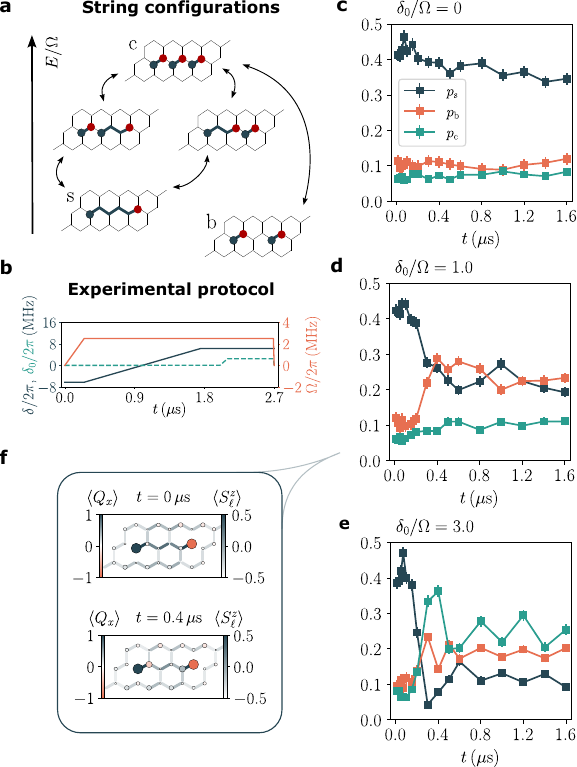}
    \caption{{\bf String breaking dynamics:} (a) Different string configurations involved in a collective process required to transition from the unbroken to the fully broken string, where we consider a system with $L_0 = 5$ and $L_1 = 3$, and static charges separated by a distance $d = 2$. The arrows indicate the direct transitions driven by $\Omega$. (b) Experimental protocol employed to adiabatically prepare a string state by ramping up the detuning globally, followed by a quench in the local detuning $\delta_0$. The initial state is prepared at $R_{\rm b} = 1.2$ and $\delta / \Omega = 2.3$, for the geometry in (a), for which, under ideal preparation, the probability of the unbroken string [s in (a)] is $p_{\rm s}(t=0)\approx 0.8$. 
    (c) -- (e) show the time-evolved probabilities obtained experimentally for the unbroken $p_{\rm s}$, broken string $p_{\rm b}$, [b in (a)] and fully charged configurations $p_{\rm c}$ [c in (a)] for different values of $\delta_0 / \Omega$. If we do not quench the system, $p_{\rm b}$ and $p_{\rm c}$ do not grow in time, while they show a fast growth followed by damped oscillations if we quench to $\delta_0 / \Omega = 1.0$ and $\delta_0 / \Omega = 3.0$, respectively.
    (f) Real-space configuration at time $t = 0$ and $t = 0.4$ $\mu$s obtained experimentally for a quench to $\delta_0 / \Omega = 1$, showing the initially unbroken and the broken strings, respectively.
    }
    \label{fig:fig4_1}
\end{figure}

Fig.~\ref{fig:fig3}(e) and (f) show the ground state phase diagram for this geometry obtained experimentally as a function of $R_{\rm b}$ and $\delta / \Omega$ (see Methods), confirming our theoretical prediction [Fig.~\ref{fig:fig1}(h)]. Within the ordered (nematic) phase outlined in Fig.~\ref{fig:fig1}(h), we observe how strings emerge between the two charges placed in the LGT vacuum, as a superposition of the six classical string configurations depicted in Fig.~\ref{fig:fig3}(a), characterized by a total string probability $p_{\rm s} = \sum_{i=0}^6 p_i$, that we obtain by measuring in the occupation basis and computing the corresponding frequencies $p_i$ for each classical string $i$ (see Methods). Fig.~\ref{fig:fig3}(c) shows how, as we increase the value of $R_{b}$ (string tension), the string probability vanishes and we instead find a non-zero probability for the fully broken string configuration $p_{\rm b}$, signaling string breaking in equilibrium. These results are consistent with the theoretical predictions depicted in Fig.~\ref{fig:fig3}(d).

Finally, in Fig.~\ref{fig:fig3}(i) and (j) we show the real-space configurations for two different sets of parameters within each region, consistent with an unbroken and a broken string. As shown in Fig.~\ref{fig:fig3}(g), we can resolve the probability corresponding to each of the six unbroken string states, where we indeed observe finite values for each of them, that we note are not all equal. This is expected since these ($2+1$)D strings do not have the same mean energy despite having the same length, as there are corrections due to the long-range Rydberg interactions depending also on the shape of the string. For example, configurations $1$ and $6$ have a lower energy compared to the others, and we indeed observe that the corresponding probabilities in our prepared states are higher for them. We note, although the obtained probabilities are reduced compared to the theory prediction due to imperfect state preparation, their qualitative features remain clearly visible. Fig.~\ref{fig:fig3}(h), shows that at a high enough string tension, these probabilities decrease, while the broken string probability acquires a finite value.  At the same time, Fig.~\ref{fig:fig3}(j) shows that the static charges are now screened by new dynamical charges produced in their vicinity, which are absent in Fig.~\ref{fig:fig3}(i).

\section*{String breaking dynamics}

We now explore how the process of string breaking occurs dynamically. To this end, we first prepare the Rydberg atom array in an initial string state following the adiabatic protocol described above, which is expected to have a lower energy than the broken string state when $2m > \sigma d$. In particular, we consider a system with $L_0 = 5$ and $L_1 = 3$ unit cells, and a pair of static charges separated by a distance $d_0 = 2$ only in the horizontal direction, corresponding to $31$ atoms [Fig.~\ref{fig:fig4_1}(a)]. We prepare the ground state of the system at $R_{\rm b} = 1.2$ and $\delta / \Omega = 2.3$, reaching initial probabilities for the unbroken and broken strings of $p_{\rm s} \approx 0.4$ and $p_{\rm b} \approx 0.1$, respectively. 

We next change the system's parameters such that $2m \approx \sigma d$, in which case the unbroken and broken string states have  similar energies, and the initial state should couple resonantly to the string broken state. For sufficiently small values of $\Omega$, one can understand this process as a higher-order coupling, where the number of off-resonant intermediate states grows linearly with $d$ [Fig.~\ref{fig:fig4_1}(a)], reminiscent of the dynamics following false-vacuum decay in spin chains~\cite{Lagnese_2021, Darbha_2024}. We thus expect the broken string probability $p_{\rm b}$ to acquire a maximum value around the resonance condition $2m \approx \sigma d$. To investigate this behaviour in our system, we quench the initial string state by adding a non-zero local detuning pattern that energetically promotes the broken string state, that is, $\delta \rightarrow \delta - \delta_0s_\ell$ for $\ell$ corresponding to $\ket{g}$ atoms in that state [Fig.~\ref{fig:fig4_1}(b) and Methods / Extended Data Fig.\ref{fig:fig5}]. By measuring the time-evolved probabilities $p_{\rm s}$ and $p_{\rm b}$ after the quench for different values of $\delta_0 / \Omega$, we can explore the resonance condition because the bare string tension acquires an extra linear contribution, $\sigma \rightarrow \sigma(R_{\rm b}) + \delta_0$. We expect that, after a certain time $t^*$, $p_{\rm b}(t \gtrsim t^*)$ develops a peak at a value $\delta^*_0 / \Omega$ for which $2m \approx \left(\sigma(R_{\rm b}) + \delta^*_0\right)d$.

Figures~\ref{fig:fig4_1}(c)--(e) show how the initial string decays in time as we quench to different values of $\delta_0 / \Omega$. For $\delta_0 / \Omega = 0$, we observe a slow decay of $p_{\rm s}$ at longer times, that we attribute to decoherence (see Methods). When we quench at  $\delta_0 / \Omega = 1.0$, however, we observe a strong decay at shorter times, as well as a simultaneous growth of $p_b$.  In Fig.~\ref{fig:fig4_1}(f), we show the real-space configuration at the initial time after state preparation and at a later time after a quench to $\delta / \Omega = 1.0$, showing how the string breaks dynamically.  Finally, for a quench at $\delta_0 / \Omega = 3.0$, we observe how a strong growth in the probability $p_{\rm c}$, associated to a state full of charges along the string location [Fig.~\ref{fig:fig4_1}(a)], while $p_{\rm s}$ quickly decays. Extended Data Fig. \ref{fig:figm3} (see Methods) shows the dynamics of probabilities after these quenches for other relevant states, where we observe how the populations for the intermediate states depicted in [Fig.~\ref{fig:fig4_1}(a)] remain low during the evolution, confirming our picture of string breaking as a high-order process.

\begin{figure}
    \centering
    \includegraphics[width=1.0\linewidth]{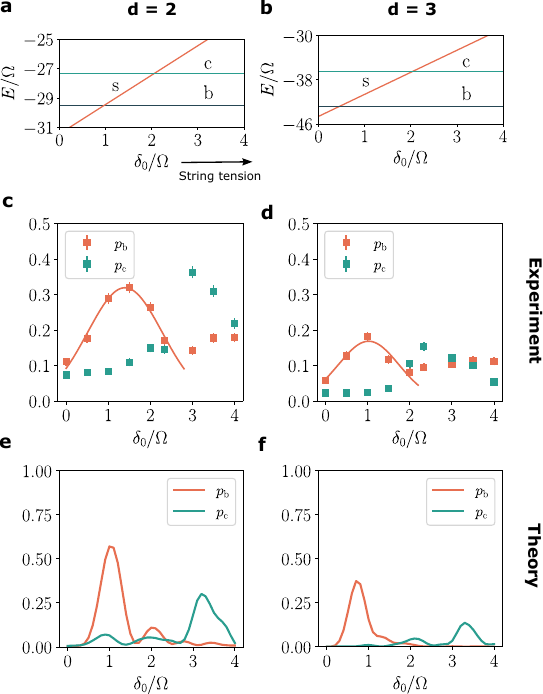}
    \caption{{\bf Many-body spectroscopy of string breaking resonances:} (a) Expectation value of the Rydberg Hamiltonian $E=\langle H \rangle$ for the product states shown in Fig.~\ref{fig:fig4_1}(a) as a function of the local detuning $\delta_0 / \Omega$, where we take $\delta / \Omega = 2.3$ and $R_{\rm b} = 1.2$. (b) Same but for a larger system size with distance $d = 3$ between the static charges. (c) Probabilities for the broken string $p_{\rm b}$  and the state full of charges $p_{\rm c}$ obtained experimentally at a fixed time $t = 0.4$ $\mu$s after the quench as a function of $\delta_0/\Omega$. The solid line corresponds to a fit of the first $6$ data points to a Gaussian function [suggested by the theoretical prediction in (e)], from which we extract a peak in $p_{\rm b}$ at $\delta_0/\Omega = 1.4(1)$. We also observe a peak in $p_{\rm c}$ of a more complicated shape. (e). Same quantities calculated numerically, showing again two peaks. For $p_{\rm b}$ the resonance is located at $\delta_0/\Omega = 1.0$, consistent with the crossing point between the states b and s shown in (a), and we attribute the difference with the experimental results to an imperfect state preparation. (d) and (f) show the the same quantities for the the case $d = 3$, where both theory and experiment present a displacement in the location of the peak as expected from the crossing point in (b). From the experimental fit using the first $5$ data points we obtain $\delta_0/\Omega = 1.0(1)$, and the numerics predict $\delta_0/\Omega = 0.7$.
    }
    \label{fig:fig4_2}
\end{figure}

We next carry out in more detail the spectroscopic measurement of the string breaking resonances. As shown in Figure~\ref{fig:fig4_2}(c), the value of $p_{\rm b}$ at a fixed time displays a peak at $\delta_0/\Omega = 1.4(1)$, within a linewidth of our numerical simulations [Fig.~\ref{fig:fig4_2}(e)], which are consistent with the point where the energies of the classical string and the broken string states cross [Figure~\ref{fig:fig4_2}(a)]. If we scan the quench parameter $\delta_0 / \Omega$ to larger values, the energy of the unbroken string  becomes resonant with the fully charged configuration [Fig.~\ref{fig:fig4_2}(a)], and we indeed observe a peak also in $p_{\rm c}$. In this case, other intermediate states have similar energies as these two states, which are thus less separated by an energy gap than in the previous case, explaining the displacement of the peak with respect to the classical level crossing [Fig.~\ref{fig:fig4_2}(a)]. Nevertheless, the peak observed experimentally, corresponding to a second string breaking resonance, is still consistent with our numerical prediction [Fig.~\ref{fig:fig4_2}(e)].

Finally, in Fig.~\ref{fig:fig4_2}(d) and Fig.~\ref{fig:fig4_2}(f), we show the results obtained after applying the same protocol to a larger system with $L_0 = 6$, $L_1 = 3$, and two static charges displaced a distance $d = 3$, for a total of $39$ atoms. The experimental data and the numerics both show a shift of the resonance peak to lower values of $\delta_0$ by 30\%. Interestingly, the the relative shift of the classical energy crossing $2m\approx\sigma d$ shown in Fig.~\ref{fig:fig4_2}(b) is a factor of 2 larger. The agreement in this observed distance dependence between the experiment and the numerics, and their deviation from the classical model, illustrate the necessity of quantum simulation to understand this quintessential feature of confinementment physics and of string breaking for larger and larger system sizes of the LGT, where numerics will no longer be able to keep up with the computational complexity.

To that end, the differences to the numerics in the absolute positions of the resonances and their widths are important to understand, and we attribute them to Rydberg blockade violations and thermal atomic motion, as described in Methods and in Extended Data Fig. \ref{fig:fig_resonance_width}

\section*{Discussion and Outlook}

Our results demonstrate the feasibility of using programmable quantum simulators to investigate both equilibrium and non-equilibrium string breaking phenomena, relevant in high-energy and condensed matter physics. While our experimental results are in good qualitative agreement with theoretical predictions in the equilibrium case and push the limits of reliable numerical simulations for non-equilibrium dynamics (Methods),  the present observations are limited by imperfections
in many-body state preparation, by Rydberg blockade violations and by decoherence arising from atomic thermal motion,  decay and dephasing  
(see Methods). The string state preparation prior to quench can be significantly enhanced via the quasi-adiabatic protocol assisted by the local detuning patterns to prepare desired strings (Methods), as demonstrated in Fig. \ref{fig:fig1}(e,f) and Extended Data Fig. \ref{fig:figm1}. %
The effects of thermal motion on both the spread of the Rydberg interaction energies and on atomic decoherence processes  could be mitigated by additional cooling and optimized trapping  (see Methods). These improvements should open the door to more extensive studies of the dynamics of superpositions of multiple strings, and to studies of the string breaking phenomenon at a larger scale.

This work lays the groundwork for new experiments, such as the study of analogs of meson scattering in (2+1)D. Moreover, our analog quantum simulation protocols could be combined with Floquet engineering to enhance multi-body plaquette interactions~\cite{Feldmeier_2024}, enabling the simulation of strong string fluctuations in ($2+1$)D, or extended to analog-digital protocols to simulate non-abelian LGTs~\cite{Gonzalez-Cuadra_2022, Zache_2023_1, Maskara_2023}. The use of such techniques to probe deconfined spin liquid~\cite{Verresen_2021,PhysRevLett.130.043601,Semeghini_2021} phases can also be explored. In a broader context the present studies provide an experimental perspective on quantum many-body systems with local constraints, including generalized lattice gauge theories such as string-net models~\cite{Levin_2005}. Finally, we emphasize that analog quantum simulations with Rydberg atom arrays scale to large particle numbers, thus allowing to solve both equilibrium and non-equilibrium dynamics in gauge theories in regimes that are classically inaccessible~\cite{Bauer2023}.

{\em Note on related work.} During the preparation of this manuscript, we became aware of two related, and complementary, studies of string breaking dynamics: on a trapped ion quantum simulator of (1+1)D LGT \cite{De2024} and on a superconducting qubit digital quantum computer implementing a (2+1)D LGT \cite{Google_2024}.

\section*{Acknowledgements}

We thank Hannes Pichler, Erez Zohar, Rhine Samajdar and Giulia Semeghini  for valuable discussions.
The Innsbruck team was supported by the European Union’s Horizon Europe research and innovation program under Grant Agreement No. 101113690 (PASQuanS2.1).  The experimental work was supported by the DARPA ONISQ program (grant number W911NF2010021) and the DARPA-STTR award (Award No. 140D0422C0035). Work at Harvard was supported by the US Department of Energy (DOE Quantum Systems Accelerator Center, grant number DE-AC02-05CH11231, and DE-SC0021013). 
The QuEra team in addition acknowledges the support of Amazon Braket in developing and validating the local detuning capability on Aquila by providing machine time and helpful discussions with P\'eter K\'om\'ar, Mao Lin and Daniela Becker.

\section*{Author contributions}
D.G.-C., T.V.Z and P.Z. developed the idea of studying the (2+1)D LGT with confinement in Rydberg atom arrays. D.G.-C., T.V.Z, B.B., M.K., A.L., F.L., S.W., A.K., M.D.L. and A.B. proposed specific experiments in this study. M.H., B.B, S.H.C., A.L. and A.B. developed the local detuning control necessary for the experiments. M.H., B.B. and A.B. performed the experiments and took the data. D.G.-C., M.H., T.V.Z and A.B. analyzed the data. D.G.-C. and T.V.Z. performed numerical simulations. M.D.L., P.Z. and A.B. guided the work and managed the resources. All authors discussed the results and contributed to writing or reviewing the manuscript.

\section*{Competing interests}
M.H., B.B., M.K., A.L., S.H.C., F.L., S.W., A.K., M.D.L. and A.B. are shareholders of, and M.H., M.K., A.L., S.H.C., F.L., S.W., A.K. and A.B. are also employees of QuEra Computing Inc.

\bibliographystyle{plain}
\bibliography{bibliography}

\setcounter{figure}{0}

\renewcommand{\figurename}{\textbf{Extended Data Fig.}}
\renewcommand{\theHfigure}{Extended Data\thefigure}

\section*{Methods}

\subsection*{Experimental apparatus and protocols}

The (2+1)D lattice gauge theory phenomena in this manuscript were observed in the programmable analog quantum simulator Aquila developed and operated by QuEra Computing Inc ~\cite{Wurtz_2023}. Aquila is based on neutral-atom arrays in optical tweezers that can be arranged in 2D and driven to the Rydberg state, through which the atoms interact. Coherent system evolution is described by the time-dependent Hamiltonian below \eqref{eq:Rydberg_H_exp},

\begin{equation}
\label{eq:Rydberg_H_exp}
H_{\rm exp}(t) = H_{\rm ZZ} + H_{\rm X}(t) + H_{\rm Z}(t),
\end{equation}
with
\begin{equation}
\begin{aligned}
& H_{\rm ZZ} = \frac{1}{2}\sum_{(\ell, \ell^\prime)} \frac{C_6}{\|{\bf x}_\ell - {\bf x}_{\ell^\prime}\|^6} n_\ell n_{\ell^\prime} \\
& H_{\rm X}(t) = \frac{\Omega(t)}{2}\sum_\ell\big(\sigma_\ell^+\exp{(i\phi(t))} + {\rm H.c.} \big) \\
& H_{\rm Z}(t) = - \sum_\ell \big(\delta(t) - c_\ell\delta_0(t)\big) n_\ell,
\end{aligned}
\end{equation}
where $n_\ell = \ket{r}_\ell\bra{r}$ and $\sigma_\ell^+ = \ket{r}_\ell\bra{g}$.

The Rydberg interaction term $H_{\rm ZZ}$ is static and can be programmed by specifying the positions $\{{\bf x}_\ell\}$ of up to 256 atoms in a two-dimensional area of 75${\rm \mu m}$ by 128${\rm \mu m}$. To implement the LGT on the hexagonal lattice in this work, the atoms (indexed by {$\ell$) are placed at the nodes of a Kagome lattice corresponding to the links of the hexagonal lattice, with different lattice constants $a$ utilized to tune the dimensionless LGT parameter $R_b/a$ (redefined $R_b/a \rightarrow R_b$ in the main text). The choice of the Rydberg state $\ket{r} \equiv \ket{^{70}S_{\rm 1/2}}$ of the Rb-87 atoms used in Aquila sets the scale for the interaction term via the coefficient value $C_6/2\pi=862,690 {\rm \ MHz \cdot\mu m^6}$ (redefined $C_6/a^6 \rightarrow C_6$ in the main text).

The $H_{\rm X}(t)$ and $H_{\rm Z}(t)$ terms are implemented by a global two-photon laser drive between the ground state $\ket{g}$ and the Rydberg state $\ket{r}$. The amplitude, frequency and phase of the laser drive are fully programmable, enabling the control over the corresponding time-dependent Hamiltonian parameters: Rabi frequency $\Omega(t)$, global laser detuning from ground-Rydberg resonance $\delta(t)$, and phase $\phi(t)$. In the experiments here, direct phase control is unused, and the Rabi frequency amplitude is always a pulse with peak value of $\Omega_{\rm max}/2\pi = 2.5 \ {\rm MHz}$, which sets the resonant blockade radius $R_b = (C_6/\Omega_{\rm max})^{1/6}$ = 8.4 ${\rm \mu m}$.

The local component of the detuning $c_\ell\delta_0(t)$ is implemented via a local AC Stark shift of the ground-Rydberg transition relative to the global laser drive, applied using an array of off-resonant laser beams with programmable static positions (overlapped with the positions of desired atoms), and programmable static intensities (controlled via the grayscale coefficients $\{c_l\} \in [0,1]$). In addition, the global scale $\delta_0(t)$  of the local detuning pattern can be controlled in time during the evolution.

\begin{figure}
    \centering
    \includegraphics[width=1\linewidth]{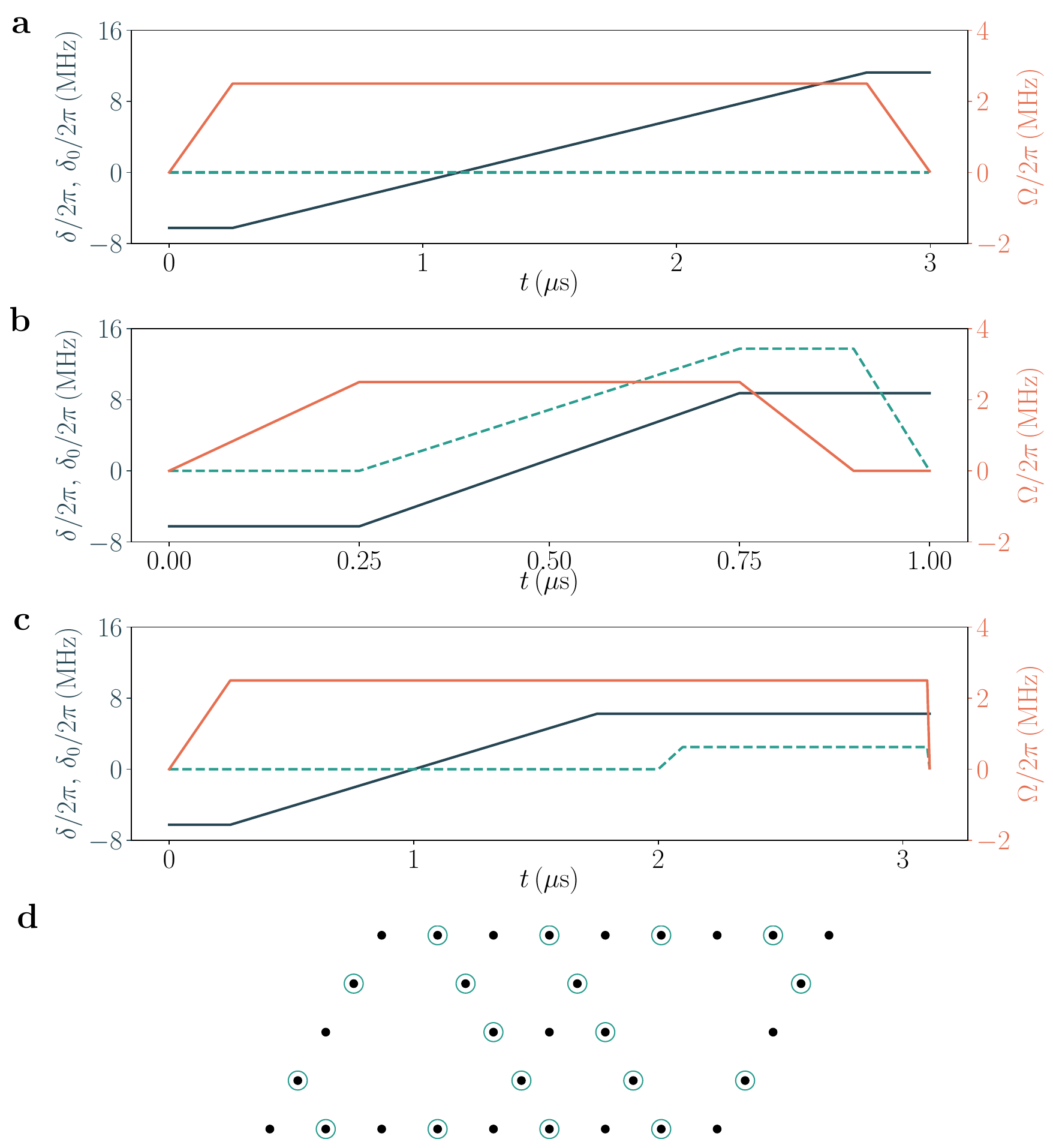}
    \caption{{\bf Experimental Hamiltonian evolution protocols} (a) Quasi-adiabatic state preparation used to obtain the ground state of the Rydberg Hamiltonian~\eqref{eq:Rydberg_H}, where the local detuning (dashed tuquoise line) remains at zero throughout the sweep. (b) Quasi-adiabatic state preparation with applied local detuning $\delta_0(t)$ (dashed tuquoise line), which strongly shifts the atoms off resonance, ensuring they remain in their ground states throughout the global detuning sweep. By applying the appropriate local detuning pattern (e.g. the one shown in (d)), one can selectively prepare either one of the string states or the broken string state. 
(c) Quasi-adiabatic state preparation followed by a quench in local detuning $\delta_0$ (dashed tuquoise line) that tensions the initially prepared string, such that the energies of the broken and unbroken string configurations become comparable. (d) For the string breaking studies with $d=2$ charge separation, the local detuning pattern applied is shown in open tuquoise circles over the atom geometry. This same detuning pattern can be used to initially prepare the broken string state via the protocol described in (b).
     }
    \label{fig:fig5}
\end{figure}

Three Hamiltonian evolution protocols are utilized in this work, starting with all atoms in the defined arrangement and in the ground state $\ket{g}$: preparation of the ground states in the confined phase via a simple quasi-adiabatic global detuning sweep (Extended Data Fig. \ref{fig:fig5}(a)), preparation of non-equilibrium states in the confined phase via the quasi-adiabatic global detuning sweep assisted by a local detuning pattern  (Extended Data Fig. \ref{fig:fig5}(b)), and a quench protocol that starts with ground state preparation via the quasi-adiabatic global detuning sweep, followed by a sudden application of a local detuning pattern that tensions the string (Extended Data Fig. \ref{fig:fig5}(c)).

For the simple global detuning sweep, the global detuning starts at a large negative value $ \delta(t=0)/2\pi = -6.25 {\rm MHz}$ while the Rabi frequency amplitude is ramped up from 0 to the maximum value of $\Omega/2\pi = 2.5 {\rm MHz}$ and kept constant while the global detuning is swept linearly through 0 to a final value $\delta/\Omega$ which is varied to study the phase diagram in Fig.~\ref{fig:fig3} (x-axis of the phase diagram). The total time for the Hamiltonian evolution is kept at $3\mu s$ (with $2.5\mu s$ allocated to the linear sweep) to stay well within the single-atom coherence times of the system, which are described in the next section of the Methods (Extended Data Fig. \ref{fig:figT2}). We quantify the adiabaticity of this protocol by $R_0^{-1}d\delta/dt$, which specifies the dimensionless sweep rate of the global detuning in units of the adiabatic criterion rate $R_0=\Omega^2/2\pi$. From the minimum value of $\delta$ to its maximum value in mapping the phase diagram in Fig. \ref{fig:fig3}, the dimensionless sweep rate varies in the range $0.56 < R_0^{-1}d\delta/dt < 1.2$, and is equal to 0.99 for the 1D data slice shown in Fig. \ref{fig:fig3}(c). Because $R_0^{-1}d\delta/dt << 1$ is not satisfied, we refer to the protocol as quasi-adiabatic in this work; the imperfect adiabaticity is the main factor that contributes to the reduced probability of actually preparing the ground state of the system at the target parameters, and limits our ability to isolate the dynamics of that state alone in the results presented in Fig.~\ref{fig:fig4_1} and Fig.~\ref{fig:fig4_2}.

For the second protocol, the quasi-adiabatic global detuning sweep described in the first protocol is assisted by a sweep of a local detuning pattern applied to the atoms that should remain in the ground state for the target product state of the system ($c_l=1$ for those $\ket{g}$ atoms and $c_l=0$ for the rest $\ket{r}$ atoms). The time-dependent $\delta_0(t)$ ensures that the total detuning for those atoms remains negative as at the start of the protocol $(\delta(t)-\delta(0))/\Omega < -2.5 $, to prevent their excitation to the Rydberg state. To return to the Hamiltonian of the system describing the LGT, $\delta_0$ is abruptly returned to 0; after that, programmable Hamiltonian evolution can proceed starting with the newly prepared non-equilibrium state of the system. Extended Data Fig. \ref{fig:figm1} shows experimental results of preparing each of the 6 strings (indexed 1-6 in Fig. \ref{fig:fig3}(a)) in our (2+1)D LGT with static charges displaced by d=2 in both dimensions. Fig. \ref{fig:fig1}(g) showcases that result for string 3, and Fig. \ref{fig:fig1}(f) shows the result for preparing a perfectly broken string state. 

\begin{figure*}
    \centering
    \includegraphics[width=0.9\linewidth]{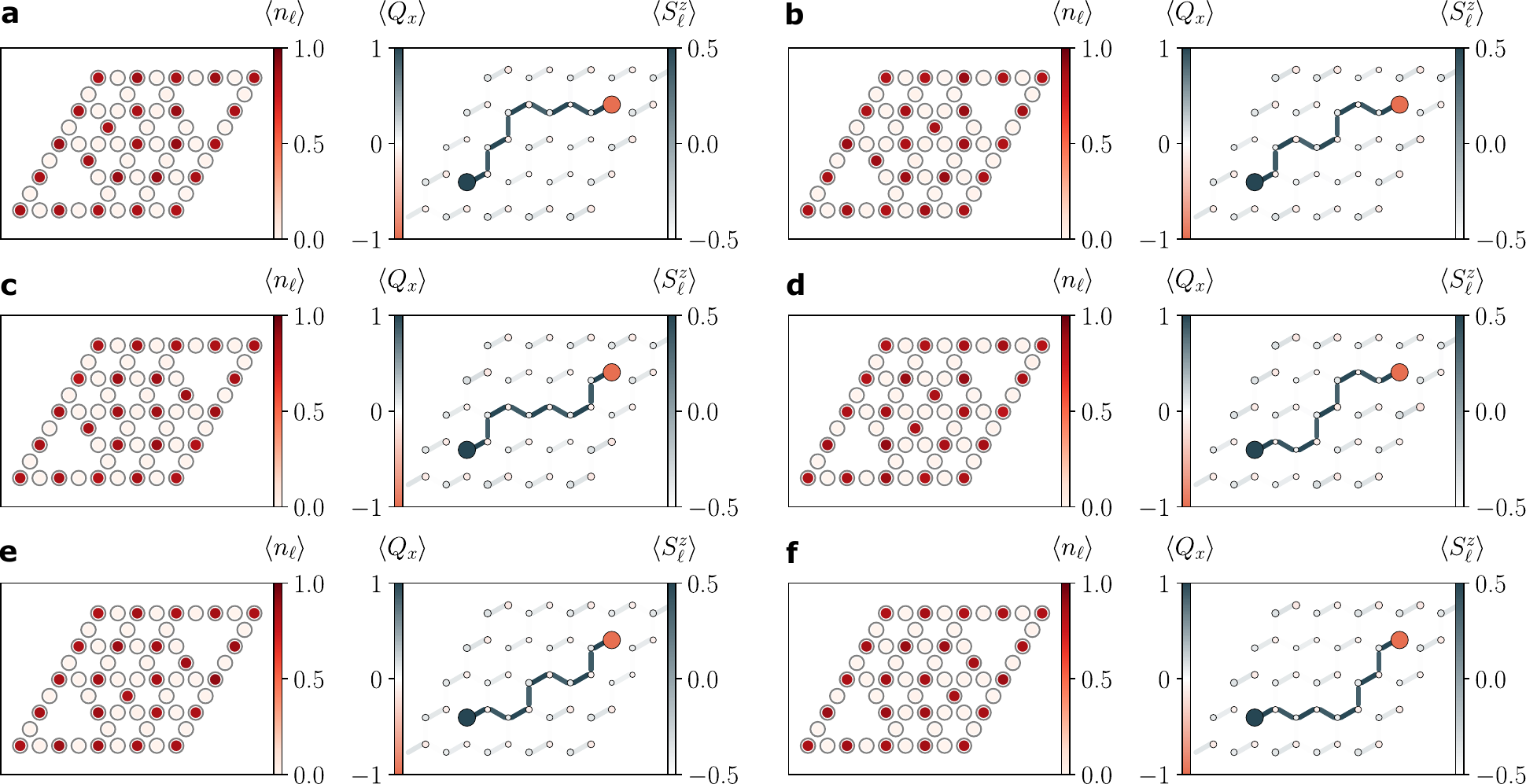}
    \caption{{\bf Experimentally prepared (2+1)D strings:} (a) -- (f) Classical string states (1) -- (6) prepared in the Rydberg atom array using the quasi-adiabatic state preparation protocol assisted by local detuning patterns. The real-space average Rydberg occupation results are represented on the left, while the extracted corresponding LGT observables are represented on the right.
    }
    \label{fig:figm1}
\end{figure*}

To study string breaking dynamics by quenching the string state with an energy penalty relative to the broken string state, we use the following protocol. We start with the quasi-adiabatic global detuning sweep described in the first protocol to prepare the string state, then quench with a local detuning pattern applied to the ground-state atoms corresponding to an ideal broken string product state ($c_l=1$ for the $\ket{g}$ atoms of a broken string). The quench is an abrupt application of $\delta_0$ that remains constant for a variable time used to time-resolve the dynamics, and this value of $\delta_0$ is also scanned to observe the resonance between the string and the broken string states. Note that to allocate up to 1.6 $\mu s$ for observing quench dynamics, the duration of the linear sweep in the quasi-adiabatic state preparation step is reduced from $2.5 \mu s$ to $1.5 \mu s$, which for the $\delta/\Omega =2.3$ used in Fig. \ref{fig:fig4_1} and Fig. \ref{fig:fig4_2} results in $R_0^{-1}d\delta/dt = 1.3$, further limiting state preparation and the isolation of the desired state in studying the subsequent dynamics.

\subsection*{Experimental imperfections and decoherence}

The control over the Hamiltonian parameters in equation \eqref{eq:Rydberg_H_exp} is subject to both systematic and random shot-to-shot errors. In addition, single-atom $T_2$ and $T_2^*$ processes intrinsic to the system lead to decoherence of the many-body dynamics which affects both the quasi-adiabatic preparation of states, and the string breaking dynamics. Finally, imperfect preparation of the individual atoms in the correct ground state, as well as imperfect ground and Rydberg state detection, produce errors in the measured outcomes even in the absence of control errors and decoherence.

Systematic spatial variations of the Rabi frequency at $\Delta\Omega_{\rm RMS}/\Omega \approx 0.02$ and of global detuning at $\Delta\delta_{\rm RMS}/2\pi \approx 150 {\rm kHz}$, are predominantly in the $y$ direction in the reference frame adopted in this manuscript. While the quasi-adiabatic protocols used in state preparation are to first order insensitive to errors in these parameters, their spatial variation may imprint spatial defects, reducing the yield of perfectly prepared states. However, this yield is currently limited by the necessary tradeoff between system coherence and non-adiabaticity effects, as discussed in the previous section on protocols.

Random static errors in the atom coordinates $\{{\bf \Delta x_l}\}$ on the order ${\rm \Delta x_{RMS},\Delta y_{RMS} \approx 0.1 \mu m}$ are compounded by random shot-to-shot position errors coming from the thermal motion of the atoms ${\rm \sigma_x, \sigma_y \approx 0.2 \mu m}$. The second component produces quasi-static noise in $H_{\rm ZZ}$ (calculated in Extended Data Fig. \ref{fig:figthermal}(a)), introducing decoherence into any many-body dynamics, whereas the first component may result in systematic shifts in measured properties of the many-body system. In the blockaded regime explored in this work, the thermal motion errors can be treated as uncorrelated single-particle $\hat{\sigma}_z$ errors on all Rydberg atoms in a given state (e.g. string or broken string), resulting from the spread of the interaction energy with all other Rydberg atoms, added in quadrature. This mean-field spread in the energies of the string and broken string states varies drastically with the chosen $R_b/a$ as shown in  Extended Data Fig. \ref{fig:figthermal}(b), but remains much smaller than the width of the many-body resonances observed in Fig. \ref{fig:fig4_2}. A much larger energy spread, coming from blockade-violating intermediate states not shown in Fig. \ref{fig:fig4_1}(a), as well as blockade-violating states accepted as string or broken string states by the spatial filtering, is likely responsible for the width of these resonance features. Experimental evidence of these effects is described in the last section of the Methods on blockade violations.

\begin{figure}
    \centering
    \includegraphics[width=1\linewidth]{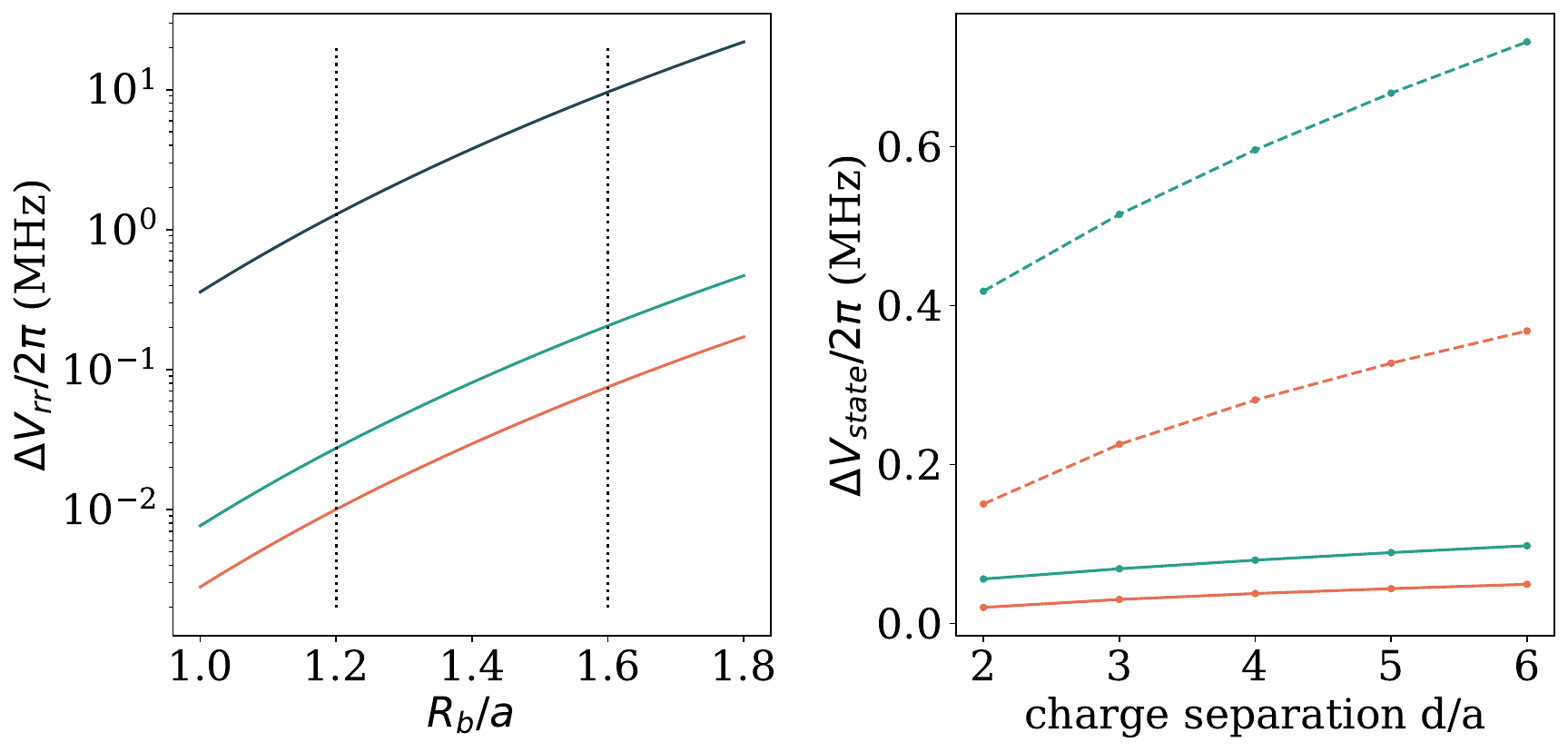}
    \caption{{\bf Decoherence due to thermal motion} (a) Thermal spread in the Rydberg-Rydberg interaction energy vs ${\rm R_b/a}$ for the different-order neighbors in the Kagome lattice (dark blue line: $\|{\bf x_1} - {\bf x_2}\| = a$, turquoise line: $\|{\bf x_1} - {\bf x_2}\| = 3^{1/2}a$, orange line: $\|{\bf x_1} - {\bf x_2}\| = 2a$), showing the dominant effect from any blockade-violating states (dark blue line). (b) Mean-field energy spreads of ideal string states (turquoise lines) and broken string states (orange lines) vs charge separation in the 1D geometries studied in Fig.\ref{fig:fig4_1} and \ref{fig:fig4_2}. Solid lines show $R_b/a=1.2$ where the dynamics data was taken, and dashed lines show $R_b/a=1.6$, deep in the confined phase. The corresponding $R_b/a$ values are marked as black dashed lines in (a).
     }
    \label{fig:figthermal}
\end{figure}

Random static  errors in the local detuning values $\{\Delta c_l\}$ on the order $\Delta c_{RMS} \approx 0.1$ do not affect either the local detuning assisted state preparation protocol, or the string breaking dynamics, as the application of the local detuning term is used to open an energy gap between degenerate configurations of Rydberg excitations.

Single-atom decoherence of the ground-Rydberg manifold is on a comparable time scale to the Hamiltonian evolution durations employed, and is a significant source of errors in state preparation, and in the many-body dynamics we observe during the string breaking quenches. To quantify this decoherence, we measure $T_2^{Rabi}$ and $T_2^{Ramsey}$ of isolated (non-interacting) atoms distributed over the same region where the Kagome lattices of atoms are arranged for the work in this manuscript (Extended Data Fig. \ref{fig:figT2}). The decaying envelope of a resonant Rabi oscillation is fit to an exponential for each atom to obtain $T_2^{Rabi}$, a proxy for single-atom decoherence processes affecting those atoms in the many-body experiments undergoing resonant coupling between $\ket{g}$ and $\ket{r}$. Similarly, the decaying envelope of a Ramsey fringe is fit to an exponential for each atom to obtain $T_2^{Ramsey}$, a proxy for single-atom decoherence processes affecting those atoms in the many-body experiments shifted from resonant coupling between $\ket{g}$ and $\ket{r}$ by the Rydberg blockade. We measure both decoherence mechanisms as a function of the magnitude of an applied local detuning to extract additional decoherence induced by this local control. This dependence for a typical atom is shown in Extended Data Fig. \ref{fig:figT2}(d). From the dependence of $T_2^{Rabi}$ on the local detuning magnitude, we extract the bare ${T_2}_0^{Rabi}$ (in the absense of local detuning), and the local-detuning-induced decoherence metric $\theta_{Rabi}$ defined as the phase angle accumulated by evolution under the given local detuning within the local-detuning-induced decoherence time ($\theta_{Rabi} = \delta_{0}\cdot {T_2}_{LD}^{Rabi}$).  The model used to extract these metrics is $1/T_2^{Rabi} = 1/{T_2}_0^{Rabi} + \delta_{0}/\theta_{Rabi}$. Averaged over the sites, these metrics yield: ${T_2}_0^{Rabi} = 3.8 \pm 0.15{\rm \mu s}$ and $\theta_{Rabi} = 2\pi\cdot(57 \pm 4)$. Similarly, the $1/T_2^{Ramsey} = 1/{T_2}_0^{Ramsey} + \delta_{0}/\theta_{Ramsey}$ model is fit to the dependence of $T_2^{Ramsey}$ on the local detuning magnitude, yielding: ${T_2}_0^{Ramsey} = 4.67 \pm 0.13{\rm \mu s}$ and $\theta_{Ramsey} = 2\pi\cdot(5.9 \pm 0.3)$, averaged over the atoms. We attribute the additional decoherence induced by local detunings to the shot-to-shot atomic position variations described earlier, sampling the intensity profile of the local detuning beam resulting in the much larger effect on $T_2^{Ramsey}$ than on $T_2^{Rabi}$ as observed in Extended Data Fig. \ref{fig:figT2}.

\begin{figure}
    \centering
    \includegraphics[width=1\linewidth]{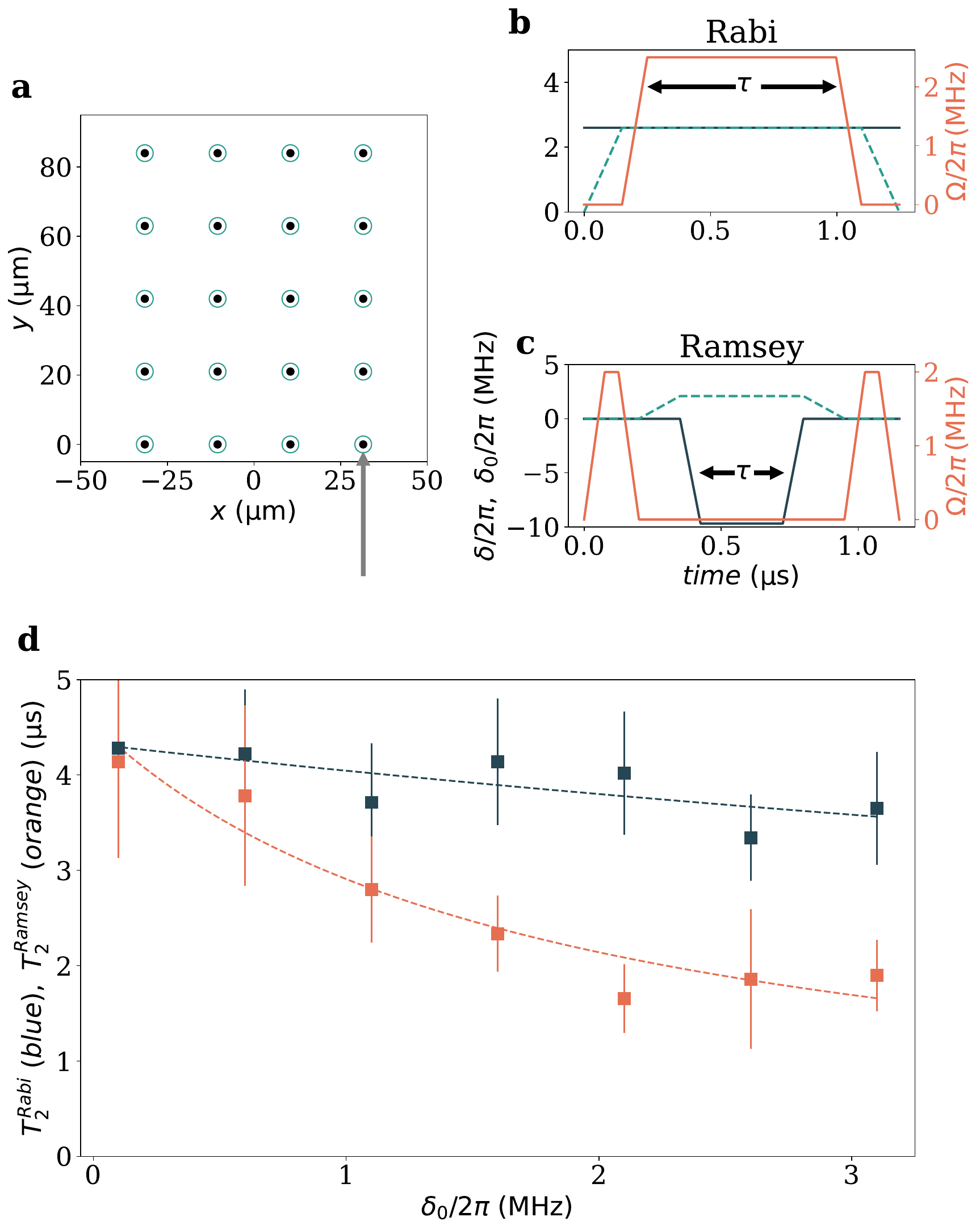}
    \caption{{\bf Single-atom coherence} (a) To measure the coherence of non-interacting atoms over the region used to arrange the atom array, a 5x4 rectangular grid of atoms is arranged instead, spaced by 21 $\mu m$. Open turquoise circles indicate atoms that have local detuning applied to them - all of them in these measurements. The gray arrow points to the atom for which $T_2^{Rabi}$ and $T_2^{Ramsey}$ results are shown in panel (d). (b) Single-atom coherence under drive is characterized by measuring resonant Rabi oscillations in the presence of local detuning (turqoise dashed line), with global detuning (dark blue line) chosen to be of equal magnitude in order to maintain resonance. The pulse duration $\tau$ is scanned to obtain Rabi oscillations in the Rydberg population, the decaying envelope of which is fit to an exponential to obtain $T_2^{Rabi}$ for each atom.  (c) Non-driven single-atom coherence is characterizeed by measuring a Ramsey fringe with local detuning (turqoise dashed line) applied during the dark time, and a large offset in the global detuning (dark blue line) applied to produce a high-frequency Ramsey fringe vs a scanned dark time $\tau$. The envelope of the fringe is fit to an exponential in order to extract $T_2^{Ramsey}$ for each atom. The Rabi frequency amplitude (orange line) shows the resonant $\pi/2$ pulses before and after. (d) Fitted values of $T_2^{Rabi}$ (dark blue) and $T_2^{Ramsey}$ (orange) vs the magnitude of the local detuning applied, for the atom highlighted with the gray arrow. The models that are fit to the data are descibed in the text.
     }
    \label{fig:figT2}
\end{figure}

Finally, we note that atom state detection fidelities can also have a substantial effect on our observations. We estimate these to be 0.95 for correct Rydberg state detection of a single atom, and 0.99 for correct ground state detection of a single atom. When characterizing state preparation, a perfectly prepared state will be obtained  with probability $0.99^{n_g} 0.95^{n_r}$ where $n_g$ and $n_r$ are the numbers of ground-state and Rydberg-state atoms, respectively, in the ideal target state in the portion of the geometry that is included in detecting the state bitstrings. 
While the state detection errors do not affect quantum dynamics,  
the observed fidelities of state preparation are lower than this state detection limit.  The additional errors are due to  non-adiabatic effects and decoherence  discussed above, which result in  the preparation of mixed states, which in turn may affect subsequent dynamics.

\subsection{Blockade violations during string breaking dynamics}

In Extended Data Fig.~\ref{fig:figm3}, we show, apart from the time-evolved string probabilities after different quenches shown in Fig.~\ref{fig:fig4_1}, the corresponding probabilities for the remaining string configurations. These include the partially-broken intermediate states shown in Fig.~\ref{fig:fig4_1}(a). We find that the relatively smaller population of these states does not grow in time. 

We also show the measured probabilities corresponding to the three atomic configurations that violate the Rydberg blockade constraint - that is, either two or three consecutive Rydberg atoms in the three-atom string. These populations are significantly lower throughout the whole evolution, but slowly increase in time, and as intermediate states in the coupling, contribute a significant thermal energy spread to the final broken string state, as per Extended Data Fig. \ref{fig:figthermal}(a). This is one source of the broadening of the experimental many-body resonance observed.

Another source of this broadening is the counting of blockade-violating states, with their large thermal energy spread as per Fig. \ref{fig:figthermal}(a), as detected broken string states when blockade violations occur on the neighbors to the three string atoms - this spatial filtering for bitstring detection purposes is explained in the next section. Extended Data Fig. \ref{fig:fig_resonance_width} confirms than when this filtering is less aggressive, the many-body resonance peak narrows thanks to post-selection, albeit at the cost of the final state probability and the signal-to-noise ratio.

\begin{figure}
    \centering
    \includegraphics[width=1.0\linewidth]{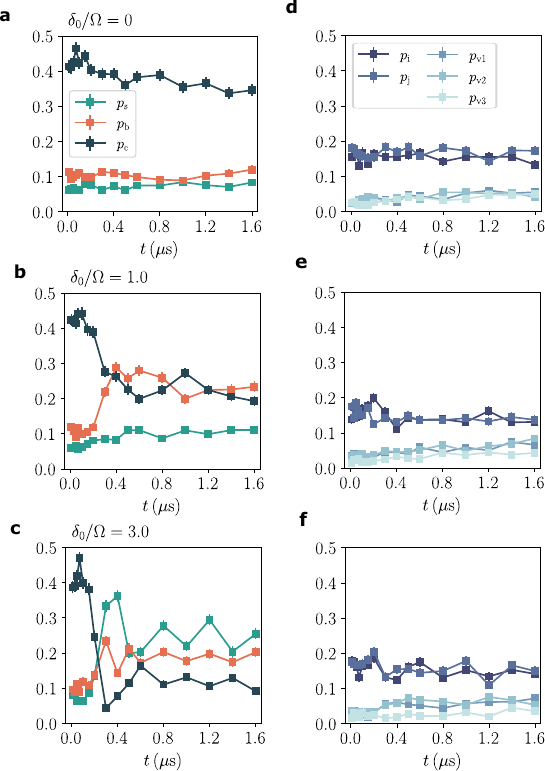}
    \caption{{\bf String probabilities and blockade violations:} (a) - (c) show the time-evolved probabilities for string configurations after a quench to different values of the local detuning $\delta_0 / \Omega$. For each of them, we also show the probabilities for the intermediate states $i$ and $j$ depicted in Fig.~\ref{fig:fig4_1}(a), as well as the three atomic configurations that violate the blockade constraint within the string ($p_{\rm v1}$, $p_{\rm v2}$ and $p_{\rm v3}$).  
    }
    \label{fig:figm3}
\end{figure}

\begin{figure}
    \centering
    \includegraphics[width=1\linewidth]{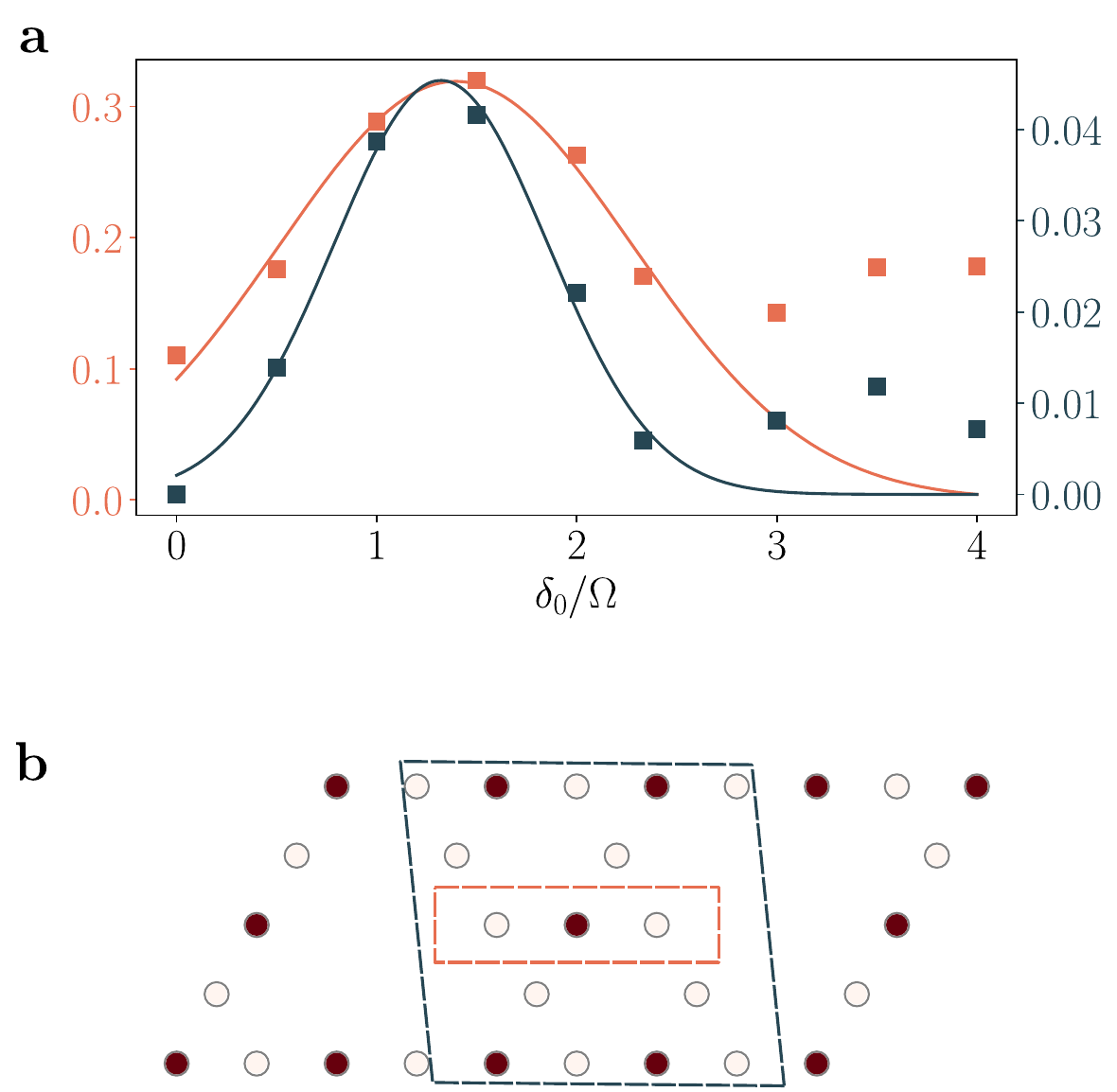}
    \caption{{\bf Broadening of string breaking resonances due to blockade violations in the final state.} (a) Broken string probability peaks under aggressive spatial filtering for bitstring detection (left axis, orange data, fitted Gaussian width $\Delta\delta_0/\Omega = 0.88(5)$), and under conservative spatial filtering (right axis, dark blue data, fitted Gaussian width $\Delta\delta_0/\Omega = 0.53(2)$). (b) For bitstring detection toward the final broken string state, the aggressive spatial filter includes all atoms in the orange box and the conservative spatial filter includes all atoms in the dark blue box. The expected configuration of Rydberg excitations in the broken string product state is shown by filled red circles.
    }
    \label{fig:fig_resonance_width}
\end{figure}

\subsection*{String probabilities}

The string probabilities are calculated by first collecting a series measurements for the atom array in the occupation basis. For each snapshot, we compare the atomic occupation within a filtered region in space with the occupation associated to each classical string configuration. For the case of static charges displaced a distance $d$ in the horizontal direction, the filtered region corresponds to the $2d-1$ atoms located between these two charges, e.g. the $3$ atoms within the string of Fig.~\ref{fig:fig4_1}(a). If the charges are displaced in both directions, the filtered region corresponds also to atoms in the region where the possible classical strings have support, e.g. the atoms belonging to the four central plaquettes in Fig.~\ref{fig:fig3}(a). The string probabilities are then defined as the frequency of occurrences of the corresponding string pattern.

To estimate the error associated with the measured probabilities, we calculate a binomial proportion confidence interval. In particular, we use a Clopper–Pearson interval with error rate $\alpha = 0.32$, such that the error bars shown in Fig.~\ref{fig:fig3}, Fig.~\ref{fig:fig4_1} and Fig.~\ref{fig:fig4_2} correspond to a confidence interval for the estimated probability of $68\%$.

\subsection*{LGT encoding}

We consider a U($1$) LGT in (2+1)D described by the following Hamiltonian,
\begin{equation}
\label{eq:QLM_H}
    H_{\rm QL} = \frac{\Omega}{2} \sum_{\langle x,y\rangle} \left[a^\dagger_x S^+_{\langle x,y \rangle}a^{\vphantom{\dagger}}_y + \text{H.c.}\right] + \frac{\delta}{2}\sum_x (-1)^{s_x} a^\dagger_x a^{\vphantom{\dagger}}_x.
\end{equation}
Here $a^{(\dagger)}_x$ denote hardcore bosonic operators defined on the sites $x$ of an hexagonal lattice, satisfying $[a^{\vphantom{\dagger}}_x, a^\dagger_{y}] = \delta_{x,y}$, with $(a^\dagger_x)^2=(a^{\vphantom{\dagger}}_x)^2=0$. $S^\pm_\ell$ and $S^z_\ell$ are spin-$1/2$ electric field operators defined on the links of the lattice, $\ell = \langle x,y \rangle$, with $[S^z_\ell, S^{\pm}_{\ell^\prime}] =\pm S^{\pm}_\ell\delta_{\ell,\ell^\prime}$ and $[S^+_\ell,S^-_{\ell^\prime}] = 2S^z_\ell\delta_{\ell,\ell^\prime}$. More precisely, $H_{\rm QL}$ corresponds to a quantum link (QL) representation~\cite{Chandrasekharan_1997, Wiese_2013} of the Abelian-Higgs model~\cite{Fradkin_1979, Gonzalez_2017}, describing quantum electrodynamics (QED) coupled to scalar matter fields. Here, the usual infinite-dimensional gauge fields are truncated and replaced by spin-$1/2$ operators, while the local U($1$) symmetry is maintained.

The first term in Eq.~\eqref{eq:QLM_H} accounts for gauge-matter interactions, while the second one corresponds to a staggered mass term, where $s_x = 0$ ($1$) for sites $x$ belonging to the A (B) sublattices shown Fig.~\ref{fig:fig1}(c). Both these terms commute with the local generator of the U($1$) gauge transformations,
\begin{equation}
    G_x = \nabla_x S^z - Q_x,
\end{equation}
where $\nabla_x S^z \equiv (-1)^{s_x}\sum_{\ell\in x} (-1)^{s_\ell} S^z_\ell$, with $s_\ell = 0$ ($1$) for links $\ell$ connecting sites that belong to the same (different) unit cell, and $Q_x = a^\dagger_x a^{\vphantom{\dagger}}_{x} - [1 - (-1)^{s_x}]/2$ are dynamical charges. Physical states $\ket{\psi}$ of this U($1$) LGT satisfy the Gauss's law constraint, $G_x\ket{\psi} = q_x\ket{\psi}$, where $q_x$ are so-called static charges. 

To encode this LGT in the Rydberg atom array, we make the following identification between operators: for links $\ell$ connecting sites within the unit cell we define $S^z_\ell = \ket{r}_\ell\!\bra{r} - 1/2$ and $S^{+}_\ell = \ket{r}_\ell\!\bra{g}$, $S^{-}_\ell = \ket{g}_\ell\!\bra{r}$, while for links connecting sites on different unit cells we take $S^z_\ell = 1/2 - \ket{r}_\ell\!\bra{r}$ and $S^{+}_\ell = \ket{g}_\ell\!\bra{r}$, $S^{-}_\ell = \ket{r}_\ell\!\bra{g}$. According to this definition, the Rydberg blockade constraint,
\begin{equation}
\sum_{\ell\in x} \ket{r}_\ell\!\bra{r} = 0, \, 1,
\end{equation}
corresponds to Gauss's law for a sector with a staggered configuration of static charges, $q_x = (-1)^{s_x}/2$.

To map $H_{\rm QL}$ to the Rydberg Hamiltonian, we first make use of the redundancy introduced by the local symmetry, and express $H_{\rm QL}$ solely in terms of electric field variables on the links by integrating out the matter degrees of freedom explicitly using Gauss's law. Using the definitions above, the resulting Hamiltonian corresponds to the Rydberg Hamiltonian restricted to the blockade subspace, $H_{\rm eff}$, and in the absence of long-range interactions. The latter can be expressed in terms of electric-field variables, resulting in $H_{\rm eff} = H_{\rm QL} 
 + H_{\rm tails}$, with
\begin{equation}
H_{\rm tails} = \frac{1}{2}\sum_{(\ell, \ell^\prime)\neq \langle\ell, \ell^\prime\rangle }V_{\ell,\ell^\prime}\left[(-1)^{s_\ell}S^z_\ell + \frac{1}{2}\right]\left[(-1)^{s_{\ell^\prime}}S^z_{\ell^\prime} + \frac{1}{2}\right].
\end{equation}

To understand confinement in the resulting model [Eq.~\eqref{eq:LGT_H}], consider first its classical limit ($\Omega = 0$). For a large mass $m_0$, the vacuum state has no charges, $Q_x = 0$ $\forall x$ (staggered occupation of hardcore bosons at half filling), and electric fields $S^z_\ell = -1/2$ $\forall \ell$, corresponding to an ordered state of the atomic system [Fig.~\ref{fig:fig1}(c)]. This is realized as a maximal covering of Rydberg excitations that satisfies the Rydberg constraints while maximizing their pair-wise distances, minimizing therefore the potential energy due to the long-range interaction tails. This ordered state is indeed the ground state of the Rydberg Hamiltonian~\eqref{eq:Rydberg_H} in the regime $\delta/\Omega \gg 1$ and $1.2 \lesssim R_b \lesssim 1.8$~\cite{Samajdar_2021} [see Fig.~\ref{fig:fig3}(c)].
Defects in this pattern are thus expected to cost energy within this phase.

With respect to this vacuum state, missing Rydberg excitations correspond to pairs of charges with classical mass $2m \approx 2m_0  - 6(R_{\rm b}/2)^6\,\Omega$ renormalized by quantum fluctuations $\propto \Omega$. Line defects have a linear energy cost as they do not maximize the distance between Rydberg atoms [Fig.~\ref{fig:fig1}(c)]. In the LGT interpretation, these correspond to $S^z_\ell = 1/2$ electric-flux strings. This mechanisms induces a linear confining potential between charges, $U(d) \sim \sigma d$, with a string tension whose leading contribution in the limit $\delta / \Omega \gg 1$ is given by $\sigma_0 \equiv V(\sqrt{3}) - V(2) \propto R^6_{\mathrm{b}}\,\Omega$. Here $U(d)$ is defined as the difference between the ground-state energy of the Rydberg Hamiltonian with a pair of static charges placed at distance $d$ (measured in unit cells) and in the absence of static charges.

\subsection*{Numerical benchmarking}

To benchmark our quantum simulator, we calculate the equilibrium phase diagram and the real-time dynamics of the Rydberg atom array using tensor-network methods~\cite{Hauschild_2018}. The ground state is obtained using the density-matrix renormalization group algorithm (DMRG), where we use matrix product states (MPS) of bond dimension up to $D = 300$. The range of the Rydberg interactions are truncated to a distance $3a$, which includes up to fifth-neighbor interactions. To simulate the system's dynamics, we use the time-dependent variational principle (TDVP) algorithm, with a time step $\delta t = 0.025 / \Omega$. String probabilities are obtained by sampling from the MPS in the occupation basis, and counting the number of occurrences of each relevant string configuration.

The boundaries between the unbroken and broken string regions (belonging to the ordered phase) and the disordered phase depicted in Fig.~\ref{fig:fig1}(h) are located by calculating the points in the parameter space for which the corresponding string probabilities acquire a value larger than $0.1$.

\subsection*{String breaking phase diagram for ($1+1$)D strings}

\begin{figure}
    \centering
    \includegraphics[width=1.0\linewidth]{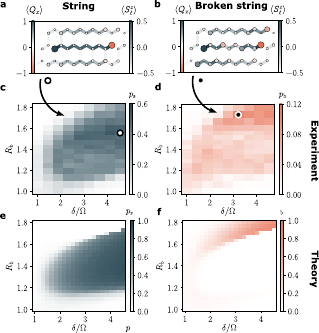}
    \caption{{\bf String breaking phase diagram for (1+1)D strings:} (a) and (b) show the real-space configuration for states with two states charges separated a distance $d = 4$, prepared with the global adiabatic protocol ending at 
    $R_{\rm b} = 1.6$, $\delta / \Omega = 4.56$ and  $R_{\rm b} = 1.7$, $\delta / \Omega = 3.22$, respectively, and consistent with an unbroken and broken strings. (c) and (d) show the unbroken ($p_{\rm s}$) and broken ($p_{\rm b}$) string probabilities, respectively, as a function of $R_{\rm b}$ and $\delta / \Omega$, obtained experimentally with the global quasi-adiabatic protocol. (e) and (f) show the corresponding theory phase diagram obtained from the ground state of the Rydberg Hamiltonian.
    }
    \label{fig:figm2}
\end{figure}

Here we consider a system with $L_0 = 7$ and $L_1 = 3$ unit cells in the horizontal and diagonal directions, respectively. We placed two static charges separated horizontally by a distance $d = 4$ by removing $6$ atoms from the array, resulting in $47$ atoms. After preparing the ground state for different values of $R_{\rm b}$ and $\delta / \Omega$, we measure in the occupation basis and calculate the probability associated to an unbroken ($p_{\rm s}$) and broken ($p_{\rm b}$) strings.

Extended Data Fig.~\ref{fig:figm2}(c) and Extended Data Fig.~\ref{fig:figm2}(d) show the experimental phase diagram for $p_{\rm s}$ and $p_{\rm b}$. We observe two regions consistent with unbroken and broken strings, signaling string breaking in the ground state of the Rydberg atom array. For a fixed and large value of $\delta / \Omega$, we expect that the mass of the charges decreases as $R_{\rm b}$ is increased, while the string tension grows, promoting string breaking. As we observed in our results, this is true also at finite values of $\delta / \Omega$, where both $m$ and $\sigma$ are renormalized by quantum fluctuations. In Extended Data Fig.~\ref{fig:figm2}(e) and Extended Data Fig.~\ref{fig:figm2}(f), we present the corresponding phase diagrams obtained numerically by calculating the ground state of the Rydberg Hamiltonian using DMRG, showing good agreement with the experimental results. Finally, Extended Data Fig.~\ref{fig:figm2}(a) and Extended Data Fig.~\ref{fig:figm2}(b) depicts the real-space configuration at different points belonging to these two regions, showing how, apart from boundary effects, the states are consistent with an unbroken and broken strings, respectively.

\end{document}